\documentclass[a4paper,12pt]{article}
\usepackage{epsfig,float}
\usepackage[left=2cm,right=2cm]{geometry}
\usepackage{graphicx}

\usepackage{epstopdf}
\usepackage{amsmath}
\usepackage{bm}
\usepackage{mathrsfs}
\usepackage{color}
\usepackage{slashed}
\usepackage{dcolumn}
\usepackage{float}

\newcommand{\be}{\begin{equation}}
\newcommand{\ee}{\end{equation}}
\newcommand{\ba}{\begin{eqnarray}}
\newcommand{\ea}{\end{eqnarray}}
\newcommand{\nn}{\nonumber}

\newcommand{\pr}{\prime}

\newcommand{\la}{\langle}
\newcommand{\ra}{\rangle}

\newcommand{\RNum}[1]{\uppercase\expandafter{\romannumeral #1\relax}}

\begin{document}

\title{The collective excitation of nuclear matter in a bosonized Landau Fermi liquid model}

\author{Bao-Xi Sun}

\maketitle

\begin{center}
Faculty of Science, Beijing University of Technology, Beijing 100124, China
\end{center}


\begin{abstract}
The collective excitation of nuclear matter is analyzed in a bosonized Landau Fermi liquid model. When the nonlinear self-interacting terms of scalar mesons are included in Walecka model, the collective excitation energy of nuclear matter can be obtained self-consistently, and the calculation results are consistent with the corresponding experimental data of the nucleus ${}^{208}Pb$ when the quantum number of the orientation of the total spin $m$ is zero.
The cases with the nonzero $m$ values are also studied, and it is found that the collective excitation energy of nuclear matter decreases with the absolute value of $m$ increasing when the total spin is conserved.
Moreover, four kinds of collective excitation modes of nuclear matter are discussed when the isospin and spin of nucleons are taken into account.
The direct interaction between two nucleons near Fermi surface only changes the effective nucleon mass and Fermi velocity, while the exchange interaction plays an critical role in the collective excitation of nuclear matter.
\end{abstract}


\section{Introduction}
\label{sect:Introduction}

The collective excitation of nuclei had been studied in
the macroscopic and microscopic model\cite{Greiner}.
Until now, it is still
an important topic in nuclear physics.
The central energy and strength
distribution can be calculated in the framework of random phase approximation, and then
the results are compared with experimental data\cite{Zhang,Ma,Ring09,SchuckRing}.

Landau Fermi liquid model has made a great success in condensed matter
physics, and some attempts have been tried to solve relevant problems in nuclear physics
with this model. By using Landau Fermi liquid model, the compression
modulus of nuclear matter is
calculated\cite{matsui}. Moreover, this model is extended to study the collective excitation of nuclei\cite{Blaizot:1978ocy,Holzwarth:1981zz,Friman96,Song:2000cu,Kamerdzhiev:2003rd,Holt:2006ii}.
In recent years, it becomes an hot point again to study the nuclear structure in the framework of Landau Fermi liquid model\cite{Ebran:2012hi,Kolomeitsev:2016zid,Roepke:2017bad,Holt:2017uuq,Grasso:2018app,
GhazanfariMojarrad:2018acp,Friman:2019ncm}.

Actually, Landau Fermi liquid model can be regarded as a method to deal with quasi-particle systems, which implies that the particles discussed in this model are not {\it real} particles. In Landau Fermi liquid model, the ground state of  Fermi system is treated as a {\it vacuum}. When a fermion is excited and jumps above Fermi surface, a hole is left in Fermi sea, which is called {\it particle-hole excitation}. In the particle-hole excitation,
these excited particles are named as {\it quasi-particles}. If a great number of quasi-particles appear above Fermi surface, the whole Fermi system lies in a collective excitation state, and the collective excitation energy of Fermi system can be calculated in Landau Fermi liquid model. Ever since it is established, this model is successfully used to study the collective excitation of electrons in metals in  condensed matter physics.

Boltzmann equation of Fermi liquid in the two-dimensional space has been discussed in Ref.~\cite{Wen}. If the external force is neglected, and supposing the quasi-particle lifetime is long enough, the bosonized equation of motion of Fermi liquid can be obtained by integrating the quasi-particle momentum. Therefore, the collective excitation of Fermi liquid becomes a wave motion of quasi-particles in the momentum space.

The bosonized Landau Fermi liquid model is extended to study the collective excitation of nuclear matter, where the isospin and spin of nucleons are taken into account\cite{Sun2010,Sun2012}.
However, Fermi liquid function, which is related to the interaction between nucleons near Fermi surface, was constructed in the original Walecka model, where the nonlinear self-interacting terms of scalar mesons are not included in the Lagrangian density, so the effective nucleon mass has to be regarded as a parameter in order to obtain the reasonable collective excitation energies, and the whole calculation is not self-consistent.

In this work, the calculation is performed in Walecka model, where the nonlinear self-interaction terms of scalar mesons are considered, and the correct energy levels of the collective excitation of nuclear matter are obtained self-consistently by using parameters in the relativistic mean-field approximation.
Moreover, four kinds of collective excitation modes of nuclear matter are analyzed when the isospin and spin of nucleons are taken into account.
In addition, the nuclear collective excitation related to the orientation of the total spin of the nucleus is also discussed.

This manuscript is organized as follows: The theoretical framework of bosonized
Landau Fermi liquid model is evaluated in detail in Sect.~\ref{sect:Boltzmann},
and the collective excitation of nuclear matter is discussed in Sect.~\ref{sect:Evskf}. Finally, the conclusion is summarized in Sect.~\ref{sect:conclusion}.
Fermi liquid function indicates the interaction of the quasi-nucleon near
Fermi surface, which is evaluated in Appendix part.

\section{Equation of motion of the collective excitation of nuclear matter}
\label{sect:Boltzmann}

If $n_{0,\vec{k}\alpha}$ represents the nucleon number in the ground
state of nuclear matter, where $\vec{k}$ denotes the nucleon momentum and $\alpha$ stands for the isospin and spin of the nucleon, and $n_{\vec{k}\alpha}(\vec{x},t)$ represents the corresponding nucleon number in the excited state, the quasi-nucleon number with $\vec{k}$ and $\alpha$ is defined as
\be
\label{eq:1905251200}
\delta n_{\vec{k}\alpha}(\vec{x},t)=n_{\vec{k}\alpha}(\vec{x},t)
-n_{0,\vec{k}\alpha},
\ee
which is a function of the position $\vec{x}$ and time $t$.

The quasi-nucleon energy in nuclear matter can be written as
\be
\label{eq:18110109332}
\tilde{\varepsilon}_{\vec{k}\alpha}(\vec{x},t)
=\xi^*_{\vec{k}\alpha}+\frac{1}{V}\sum_{\vec{k}^\prime,\beta}
f(\vec{k},\alpha;\vec{k}^\prime,\beta)
\delta n_{\vec{k}^\prime,\beta}(\vec{x},t), \ee
where $\xi^*_{\vec{k}\alpha}$ is the energy of the
quasi-nucleon with $\vec{k}$ and $\alpha$,
and Fermi liquid function $f(\vec{k},\alpha;\vec{k}^\prime,\beta)$
stands for the interaction between two quasi-nucleons with
$\vec{k}(\vec{k}^\prime)$ and $\alpha(\beta)$ respectively.

In the spherical coordinate space, Eq.~(\ref{eq:18110109332}) takes the form of
\be
\tilde{\varepsilon}_{\vec{k}\alpha}(\vec{x},t)
=\xi^*_{\vec{k}\alpha}+\frac{1}{(2\pi)^3}\sum_{\beta} \int {k^\pr}^2
d k^\pr \int \sin \theta^\pr d \theta^\pr \int d \phi^\pr
f(k, \theta, \phi, \alpha;k^\pr,\theta^\pr, \phi^\pr,\beta)
\delta n_{\vec{k}^\prime,\beta}(\vec{x},t). \ee
Therefore,
\be \label{eq:1901271440} \frac{\partial
}{\partial \vec{x}}\tilde{\varepsilon}_{\vec{k}\alpha}(\vec{x},t)
=\frac{1}{(2\pi)^3}\sum_{\beta} \int {k^\pr}^2 d k^\pr \int \sin
\theta^\pr d \theta^\pr \int d \phi^\pr
f(k, \theta, \phi, \alpha;k^\pr,\theta^\pr, \phi^\pr,\beta)
\frac{\partial}{\partial
\vec{x}}  \delta n_{\vec{k}^\prime,\beta}(\vec{x},t). \ee
At Fermi surface,
 \ba \label{eq:1901291122}
&&\frac{\partial \tilde{\varepsilon}_{\vec{k}\alpha}} {\partial
\vec{k}}=v_F^* \hat{\vec{k}},
\nn \\
&&\frac{\partial n_{\vec{k}\alpha}}{\partial \vec{k}}
=\frac{\partial {n_0}_{\vec{k}\alpha}}{\partial \vec{k}}
+\frac{\partial \delta n_{\vec{k}\alpha}}{\partial \vec{k}} \approx
\frac{\partial {n_0}_{\vec{k}\alpha}}{\partial \vec{k}}
=-\hat{\vec{k}} \delta(|\vec{k}|-k_F), \ea
with $v_F^*$ Fermi velocity, and $\hat{\vec{k}}=\vec{k}/|k|$ the unit vector
on the direction of $\vec{k}$.

According to Hamilton principle, we obtain
\be
\frac{\partial \vec{x}}{\partial t} =\frac{\partial} {\partial
\vec{k}} \tilde{\varepsilon}_{\vec{k}\alpha}(\vec{x},t), \ee and \be
\label{eq:1811031830} \frac{\partial \vec{k}}{\partial t} =-
\frac{\partial } {\partial
\vec{x}}\tilde{\varepsilon}_{\vec{k}\alpha}(\vec{x},t). \ee
Thus Boltzmann equation of nucleons can be written as
 \ba \label{eq:1811031752} \frac{d
n}{d t}&=&\frac{\partial n}{\partial t} +\frac{\partial n}{\partial
\vec{x}} \cdot \frac{\partial \vec{x}}{\partial t} +\frac{\partial
n}{\partial \vec{k}} \cdot
\frac{\partial \vec{k}}{\partial t} \nn \\
&=&\frac{\partial n}{\partial t} +\frac{\partial n}{\partial
\vec{x}} \cdot \frac{\partial }{\partial \vec{k}}
\tilde{\varepsilon}_{\vec{k}\alpha}(\vec{x},t) -\frac{\partial
n}{\partial \vec{k}} \cdot
\frac{\partial }{\partial \vec{x}} \tilde{\varepsilon}_{\vec{k}\alpha}(\vec{x},t)\nn \\
&=&I[n], \ea
with $I[n]$ the collision term.

If the external force is taken into account,
Boltzmann equation in Eq.~(\ref{eq:1811031752}) becomes
\be \label{eq:1811031757} \frac{\partial
n}{\partial t} +\frac{\partial n}{\partial \vec{x}} \cdot
\frac{\partial }{\partial \vec{k}}
\tilde{\varepsilon}_{\vec{k}\alpha}(\vec{x},t) +\frac{\partial
n}{\partial \vec{k}} \cdot \left(\vec{F}- \frac{\partial }{\partial
\vec{x}} \tilde{\varepsilon}_{\vec{k}\alpha}(\vec{x},t)
\right)
=I[n]. \ee

In the relaxation time approximation, the collision term is inverse to the relaxation time. The magnitude of the relaxation time has the same order as the average time between two continuous collisions of a nucleon, which is proportional to the nucleon mean free path in the nucleus. Moreover, due to Pauli principle, the nucleon mean free path is large at the saturation density. For the nucleon near Fermi surface, the nucleon kinetic energy is about 40MeV, and the nucleon mean free path is about 10fm\cite{Mayugang}.
Comparing to Fermi momentum of 1.36fm$^{-1}$, the relaxation time is long enough in the nucleus.
Therefore, the collision term in Eq.~(\ref{eq:1811031757}) is eliminated in the relaxation time approximation.

According to Eq.~(\ref{eq:1905251200}),
$n_{\vec{k}\alpha}(\vec{x},t)
=n_{0,\vec{k}\alpha}+\delta n_{\vec{k}\alpha}(\vec{x},t)$, we obtain
$\frac{\partial n_{\vec{k}\alpha}}{\partial t}=\frac{\partial }{\partial t}
\delta n_{\vec{k}\alpha}$ and $\frac{\partial n_{\vec{k}\alpha}}{\partial \vec{x}} =\frac{\partial
}{\partial \vec{x}} \delta n_{\vec{k}\alpha}$,
thus Eq.~(\ref{eq:1811031757}) becomes
\be
\label{eq:1905251209}
\frac{\partial \delta n_{\vec{k}\alpha}}{\partial t} +\frac{\partial
\delta n_{\vec{k}\alpha}}{\partial \vec{x}} \cdot \frac{\partial
}{\partial \vec{k}} \tilde{\varepsilon}_{\vec{k}\alpha}(\vec{x},t)
+\frac{\partial n_{\vec{k}\alpha}}{\partial \vec{k}} \cdot
\left(\vec{F}- \frac{\partial }{\partial \vec{x}}
\tilde{\varepsilon}_{\vec{k}\alpha}(\vec{x},t)
\right)
=0. \ee

In Boltzmann equation, the gravity and the electromagnetic force are often treated as external forces. However, they can be neglected in nuclei because they are weaker than the strong interaction between nucleons. Thus the external force is treated as zero in Eq.~(\ref{eq:1905251209}).

Supposing $\vec{F}=0$, Eq.~(\ref{eq:1905251209}) takes the form of
\ba \label{eq:1901271456} &&\frac{\partial }{\partial t}\delta
n_{\vec{k}\alpha}(\vec{x},t) +\frac{\partial \delta
n_{\vec{k}\alpha}(\vec{x},t)}{\partial \vec{x}} \cdot
v_F^* \hat{\vec{k}}  \\
&+&\hat{\vec{k}} \delta(|\vec{k}|-k_F) \cdot \left(
\frac{1}{(2\pi)^3}\sum_{\beta} \int {k^\pr}^2 d k^\pr \int \sin
\theta^\pr d \theta^\pr \int d \phi^\pr
f(k, \theta, \phi, \alpha;k^\pr,\theta^\pr, \phi^\pr,\beta)
\frac{\partial}{\partial \vec{x}} \delta
n_{\vec{k}^\prime,\beta}(\vec{x},t)
\right)
 =0. \nn  \ea
In the momentum space, $\frac{1}{i}\frac{\partial}{\partial \vec{x}}
\rightarrow \vec{q}$, we obtain \ba
\label{eq:1901271459} &&i\frac{\partial }{\partial t}\delta
n_{\vec{k}\alpha}(\vec{q},t) =\left(\hat{\vec{k}} \cdot \vec{q}
\right)\left(
v_F^* \delta n_{\vec{k}\alpha}(\vec{q},t) \right.  \\
&+& \left. \delta(|\vec{k}|-k_F)
\frac{1}{(2\pi)^3}\sum_{\beta} \int {k^\pr}^2 d k^\pr \int \sin
\theta^\pr d \theta^\pr \int d \phi^\pr
f(k, \theta, \phi, \alpha;k^\pr,\theta^\pr, \phi^\pr,\beta)
\delta n_{\vec{k}^\prime,\beta}(\vec{q},t)
\right). \nn
\ea

Since the quasi-nucleon number $\delta n_{\vec{k}\alpha}(\vec{x},t)$ will be a over
complete set of variables when the collective excitation of nuclear matter is studied, it turns out that
Fermi surface displacement from Fermi ball with the radius of Fermi momentum $k_F$ is proper to describe the
collective fluctuation of Fermi liquid. In this work, the Fermi surface displacement from Fermi ball in the momentum space is closely related to the nuclear quasi-nucleon density mentioned below.

The nuclear quasi-nucleon density is defined as
\begin{equation}
\label{eq:qpden}
 \tilde{\rho}_\alpha(\theta,\phi)~=~\int \frac{k^2 dk}{(2\pi)^3}
 \delta n_{\vec{k} \alpha},
\end{equation}
with
\begin{equation}
\delta n_{\vec{k} \alpha}~=~\left\{
\begin{array}{c}
1,  \\
0, \\
-1.
\end{array}
\right.
\end{equation}
In Eq. (\ref{eq:qpden}), the label $\alpha$ indicates the isospin and spin of quasi-nucleons, i.e., $\alpha=1,2,3,4$.
The quasi-nucleon number with $\vec{k}$ and $\alpha$ can be written as the formula in Eq.~(\ref{eq:1905251200}), and the ground state nucleon number $n_{0,\vec{k} \alpha}$ equals unity for nucleon momenta below Fermi momentum and vanishes for momenta above Fermi momentum. When a nucleon-hole pair is excited, a nucleon appears above Fermi surface, which corresponds to the case of $n_{\vec{k} \alpha}=1$,
$n_{0,\vec{k} \alpha}=0$ and $\delta n_{\vec{k} \alpha}=1$. Meanwhile, a hole is left below Fermi surface, which corresponds to the case of $n_{\vec{k} \alpha}=0$,
$n_{0,\vec{k}\alpha}=1$ and $\delta n_{\vec{k}\alpha}=-1$. Therefore, the quasi-nucleon number $\delta n_{\vec{k}\alpha}$ takes values of 1, 0 and -1, respectively.  Moreover, it should be pointed out that the value of $\delta n_{\vec{k} \alpha}$ will not exert an influence on the Hamiltonian obtained finally, and corresponding eigenenergies remain unchanged.

The reduced Boltzmann equation is obtained by performing the integration
$\int \frac{k^2 dk}{(2\pi)^3}$ on both sides of Eq.~(\ref{eq:1901271459}),
 \ba
\label{eq:1901280955} &&i\frac{\partial }{\partial
t}\tilde{\rho}_\alpha(\theta,\phi;\vec{q},t)
=\left(\hat{\vec{k}} \cdot \vec{q} \right)\left(
v_F^* \tilde{\rho}_\alpha(\theta,\phi;\vec{q},t) \right.  \\
&+& \left.
\frac{k_F^2}{(2\pi)^6}\sum_{\beta} \int {k^\pr}^2 d k^\pr \int \sin
\theta^\pr d \theta^\pr \int d \phi^\pr
f(k_F, \theta, \phi, \alpha;k^\pr,\theta^\pr, \phi^\pr,\beta)
\delta n_{\vec{k}^\prime,\beta}(\vec{q},t)
\right). \nn
\ea
If the nucleon near Fermi surface on
the collective
excitation of nuclear matter is taken into account, the nucleon
momentum $k^\pr$ in Fermi liquid
function $f(k_F, \theta, \phi,
\alpha;k^\pr,\theta^\pr, \phi^\pr,\beta)$ takes the value of Fermi momentum
$k_F$. Thus Eq.~(\ref{eq:1901280955}) becomes
\ba \label{eq:1901281044} &&i\frac{\partial }{\partial
t}\tilde{\rho}_\alpha(\theta,\phi;\vec{q},t)
=\left(\hat{\vec{k}} \cdot \vec{q} \right)\left(
v_F^* \tilde{\rho}_\alpha(\theta,\phi;\vec{q},t) \right.  \\
&+& \left.
\frac{k_F^2}{(2\pi)^3}\sum_{\beta} \int \sin \theta^\pr d \theta^\pr
\int d \phi^\pr
f(k_F, \theta, \phi, \alpha;k_F,\theta^\pr, \phi^\pr,\beta)
\tilde{\rho}_\beta(\theta^\pr,\phi^\pr;\vec{q},t)
\right). \nn
\ea

Supposing $(\theta_q,\phi_q)$ denotes the direction of $\vec{q}$ in the
spherical coordinate of the momentum space,  we can obtain
\be \hat{\vec{k}} \cdot \vec{q}=
q~\left[\sin \theta \sin \theta_q \cos (\phi-\phi_q) \right] +\cos
\theta \cos \theta_q, \ee
so Eq.~(\ref{eq:1901281044}) can be written as
\begin{eqnarray}
\label{eq:liquid-theta-phi} i\frac{\partial}{\partial t}
\tilde{\rho}_\alpha(\theta,\phi,\vec{q},t)~=~q \sum_\beta \int
d\Omega^{\prime} \int d \Omega^{\prime
\prime}K(\theta,\phi;\theta^\prime,\phi^\prime)
M(\theta^\prime,\phi^\prime,\alpha;\theta^{\prime\prime},\phi^{\prime\prime},\beta)
\tilde{\rho}_\beta(\theta^{\prime\prime},\phi^{\prime
\prime},\vec{q},t),
\end{eqnarray}
with
\begin{equation}
K(\theta,\phi;\theta^\prime,\phi^\prime)~=~[\sin \theta \sin
\theta_q \cos(\phi-\phi_q)~+~\cos \theta \cos \theta_q]\frac{1}{\sin
\theta^\prime }\delta(\theta-\theta^\prime)\delta(\phi-\phi^\prime),
\end{equation}
and
\begin{equation}
\label{eq:1901291158}
M(\theta,\phi,\alpha;\theta^\prime,\phi^\prime,\beta)~=~v^\ast_F
\frac{1}{\sin \theta^\prime} \delta_{\alpha \beta}
\delta(\theta-\theta^\prime)\delta(\phi-\phi^\prime)
~+~\frac{k^2_F}{(2\pi)^3}f(k_F, \theta,\phi,\alpha; k_F,
\theta^\prime,\phi^\prime,\beta).
\end{equation}

Actually, Fermi liquid function denotes the interaction between
two quasi-nucleons near Fermi surface,
and in this work, it is evaluated from the Lagrangian density of Walecka model.
The detailed process can be found in the appendix part of this manuscript.
According to Eq.~(\ref{eq:1901241841}),
Fermi liquid function can be written as

\begin{eqnarray}
\label{eq:fermifun} f(\vec{k}_1, \alpha; \vec{k}_2,
\beta)&=&V_{eff}(0)-V_{eff}(\vec{k}_1-\vec{k}_2)~\delta_{\alpha
\beta } \nonumber  \\
&=&\left(\frac{-g^2_\sigma}
{m^2_\sigma}~+~\frac{g^2_\omega}{m^2_\omega}\right)-
\left(\frac{-g^2_\sigma}
{(\vec{k_1}-\vec{k_2})^{2}+m^2_\sigma}~+~\frac{g^2_\omega}{(\vec{k_1}-\vec{k_2})^{2}+m^2_\omega}\right)
\delta_{\alpha \beta},
\end{eqnarray}
where
\begin{equation}
\vec{k}_1~=~(k_F, \theta, \phi),~~~~\vec{k}_2~=~(k_F, \theta^\prime,
\phi^\prime),
\end{equation}
and
\begin{eqnarray}
(\vec{k_1}-\vec{k_2})^{2}&=&2 k^2_F \left\{
1~-~\left[\cos{\theta}\cos{\theta^\prime}~+~\sin{\theta}\sin{\theta^\prime}\cos{(\phi-\phi^\prime)}\right]\right\}
\nonumber \\
&=&2 k^2_F \left( 1~-~ \hat{\vec{k}}_1 \cdot \hat{\vec{k}}_2
\right).
\end{eqnarray}

In Fermi liquid function in Eq.~(\ref{eq:fermifun}), the direct
interaction between nucleons gives a contribution to
the ground energy of nuclear matter, and is not relevant to the
excitation of nucleon-hole pairs directly. In the relativistic mean-field
approximation, Fermi energy of nucleons takes the form of
\begin{equation}
\label{eq:202007311120}
\varepsilon^\ast_F \simeq M_N +
\frac{k_F^2}{2M^\ast_N}+(-\frac{g_\sigma^2}{m_\sigma^2}+\frac{g_\omega^2}{m_\omega^2})
\sum_\gamma \frac{k_F^3}{6\pi^2},
\end{equation}
where the summation runs over the isospin and spin
of nucleon in nuclear matter. Correspondingly, the nucleon Fermi velocity
$v^\ast_F$ can be written as
\begin{equation}
\label{eq:velocity} v^\ast_F= \frac{\partial
\varepsilon^\ast_F}{\partial
k_F}=\frac{k_F}{M^\ast_N}+(-\frac{g_\sigma^2}{m_\sigma^2}+\frac{g_\omega^2}{m_\omega^2})
\frac{k_F^2}{2\pi^2},
\end{equation}
where the second term comes from the direct interaction term
of Fermi liquid function in Eq.~(\ref{eq:fermifun}). Therefore, only
the exchanging interaction between nucleons is essential to excite nucleon-hole pairs in nuclear matter.
Especially, if the isospin and spin of two interacting nucleons near Fermi surface are the same as each other, the exchange interaction(the second term in Fermi liquid function) will play a role in the collective excitation of nuclear matter.
For the ground state of nuclear matter, Fermi momentum $k_F$ is a constant, and the nucleon distribution in momentum space forms a Fermi ball with the radius of $k_F$.  The quasi-nucleon density represents the surface displacement from Fermi ball at the direction of $(\theta, \phi)$, so it can be expanded in spherical harmonics, i.e.,
\begin{equation}
\label{eq:rho_l}
\tilde{\rho}_\alpha(\theta,\phi,\vec{q},t)~=~\sum_{l,m}\tilde{\rho}_\alpha(l,m,\vec{q},t)Y^\ast_{l,m}(\theta,\phi).
\end{equation}
Similarly, $K(\theta,\phi;\theta^\prime,\phi^\prime)$
and $M(\theta^\prime,\phi^\prime,\alpha;\theta^{\prime\prime},\phi^{\prime\prime},\beta)$
can be expanded as
\begin{equation}
K(\theta,\phi;\theta^\prime,\phi^\prime)~=~\sum_{l,m,l^\prime,m^\prime}
K(l,m;l^\prime,m^\prime)Y^\ast_{l,m}(\theta,\phi)~Y_{l^\prime,m^\prime}(\theta^\prime,\phi^\prime),
\end{equation}
and
\begin{equation}
M(\theta^\prime,\phi^\prime,\alpha;\theta^{\prime\prime},\phi^{\prime\prime},\beta)
~=~\sum_{l_1,m_1,l_2,m_2} M(l_1,m_1,\alpha; l_2, m_2,
\beta)Y^\ast_{l_1,m_1}(\theta^\prime,\phi^\prime)~Y_{l_2,m_2}(\theta^{\prime\prime},\phi^{\prime\prime}),
\end{equation}
respectively.
Therefore, the equation of motion of Fermi liquid takes the form of
\begin{eqnarray}
\label{eq:liquid-theta-phi2}    i\frac{\partial}{\partial t}
\tilde{\rho}_\alpha(l,m,\vec{q},t) ~=~ q \sum_\beta
\sum_{l^\prime,m^\prime}
\sum_{l^{\prime\prime},m^{\prime\prime}}K(l,m;l^\prime,m^\prime)
M(l^\prime,m^\prime,\alpha; l^{\prime\prime}, m^{\prime\prime},
\beta)
\tilde{\rho}_\beta(l^{\prime\prime},m^{\prime\prime},\vec{q},t).
\end{eqnarray}

Since the collective excitation of nuclear matter is independent on
the direction of $\vec{q}$, we can choose
$\theta_q~=~0$ and $\phi_q~=~0$, and the function
$K(\theta,\phi;\theta^\prime,\phi^\prime)$ can be written as
\begin{equation}
K(\theta,\phi;\theta^\prime,\phi^\prime)~=~\frac{\cos \theta}{\sin
\theta^\prime }\delta(\theta-\theta^\prime)\delta(\phi-\phi^\prime).
\end{equation}

In the spherical harmonics of the momentum space,
\begin{eqnarray}
\label{eq:1901302047} K(l,m;l^\prime,m^\prime)&=&\int \frac{\cos
\theta}{\sin \theta^\prime
}\delta(\theta-\theta^\prime)\delta(\phi-\phi^\prime)
Y_{l,m}(\theta,\phi)~Y^\ast_{l^\prime,m^\prime}(\theta^\prime,\phi^\prime)
\sin \theta^\prime d \theta^\prime d \phi^\prime \sin \theta d
\theta d \phi \nonumber \\
&=& \left(a_{lm} \delta_{l+1,l^\prime} ~+~a_{l-1,m}
\delta_{l-1,l^\prime} \right)\delta_{m,m^\prime},
\end{eqnarray}
with
\begin{equation}
\label{eq:1903171236}
a_{l,m}~=~\sqrt{\frac{(l+1)^2-m^2}{(2l+1)(2l+3)}}.
\end{equation}
In order to obtain Eq.~(\ref{eq:1901302047}), the equation
$ \cos \theta
Y_{l,m}(\theta,\phi)=a_{l,m}Y_{l+1,m}(\theta,\phi)
+a_{l-1,m}Y_{l-1,m}(\theta,\phi)
$
is used.
Moreover,
\begin{eqnarray}
\label{eq:1909250745}
 M(l_1,m_1,\alpha; l_2, m_2,
\beta)&=&\int [v^\ast_F \frac{1}{\sin \theta^\prime} \delta_{\alpha
\beta} \delta(\theta-\theta^\prime)\delta(\phi-\phi^\prime)
~+~\frac{k^2_F}{(2\pi)^3}f(k_F,\theta,\phi,\alpha; k_F,
\theta^\prime,\phi^\prime,\beta)]  \nonumber \\
&&Y_{l_1,m_1}(\theta,\phi)~Y^\ast_{l_2,m_2}(\theta^{\prime},\phi^{\prime})
d \Omega d \Omega^{\prime} \nonumber \\
&=&v^\ast_F  \delta_{\alpha \beta} \delta_{l_1,l_2}
\delta_{m_1,m_2}~-~\frac{k^2_F}{(2\pi)^3}
f_F(l_1,m_1;l_2,m_2)\delta_{\alpha,\beta},
\nonumber \\
\end{eqnarray}
where Fermi liquid function $f_F(l_1,m_1;l_2,m_2)$ is only related to
the exchange interaction between nucleons.
\begin{eqnarray}
\label{eq:1909250746}
f_F(l_1,m_1;l_2,m_2)&=&\int
V_{eff}(\vec{k}_1-\vec{k}_2)
Y_{l_1,m_1}(\theta,\phi)~Y^\ast_{l_2,m_2}(\theta^{\prime},\phi^{\prime})
\sin \theta d \theta d \phi \sin \theta^\prime d \theta^\prime d
\phi^\prime \nonumber \\
&=&\int \left(\frac{-g^2_\sigma}
{(\vec{k_1}-\vec{k_2})^{2}+m^2_\sigma}~+~\frac{g^2_\omega}{(\vec{k_1}-\vec{k_2})^{2}+m^2_\omega}\right)
Y_{l_1,m_1}(\theta,\phi)~Y^\ast_{l_2,m_2}(\theta^{\prime},\phi^{\prime})
\nonumber \\ && \sin \theta d \theta d \phi \sin \theta^\prime d
\theta^\prime d
\phi^\prime \nonumber \\
&=&f_F(l_1,l_2) \delta_{l_1,l_2} \delta_{m_1,m_2}.
\end{eqnarray}
Apparently, the exchange interaction between two nucleons plays an important role
on the collective excitation of nuclear matter when $l_1=l_2$ and
$m_1=m_2$.

According to Eqs.~(\ref{eq:1901302047}), (\ref{eq:1909250745}) and (\ref{eq:1909250746}), the equation of motion
of nuclear matter in Landau Fermi liquid model can be written as
\begin{eqnarray}
\label{eq:liquid-2}    i\frac{\partial}{\partial t}
\tilde{\rho}_\alpha(l,m,\vec{q},t) &=&q \sum_{l^\prime} \left(a_{lm}
\delta_{l+1,l^\prime} ~+~a_{l-1,m} \delta_{l-1,l^\prime} \right)
\left(v^\ast_F-\frac{k^2_F}{(2\pi)^3}f_F(l^{\prime},l^{\prime
})\right)
\tilde{\rho}_\alpha(l^{\prime},m,\vec{q},t), \nonumber \\
\end{eqnarray}
which can be rewritten in the matrix form of
\begin{eqnarray}
\label{eq:matrixeq}    i\frac{\partial}{\partial t}
\tilde{\rho}_\alpha(l,m,\vec{q},t) &=&q \tilde{K} \tilde{M}
\tilde{\rho}_\alpha(l,m,\vec{q},t), \nonumber \\
\end{eqnarray}
with
 \begin{equation}
 \tilde{K}_{l,l^\prime}=\left(a_{lm} \delta_{l+1,l^\prime}
~+~a_{l-1,m} \delta_{l-1,l^\prime} \right),
\end{equation}
and
 \begin{equation}
\tilde{M}_{l^\prime,l}=\left(v^\ast_F-\frac{k^2_F}{(2\pi)^3}f_F(l^{\prime},l^{\prime
})\right)\delta_{l^\prime,l}\delta_{m^\prime,m}.
\end{equation}
Fermi liquid is unstable if one of the $f_F(l^{\prime},l^{\prime })$
is larger than $(2\pi)^3 \frac{v^\ast_F}{k_F^2}$,
Supposing
$\tilde{M}=WW^T$ and $u_\alpha=W^T\tilde{\rho}_\alpha$,
Eq.~(\ref{eq:matrixeq}) becomes
\begin{eqnarray}
\label{eq:Schrodinger}    i\frac{\partial}{\partial t}
u_\alpha(l,m,\vec{q},t) &=&q W^T \tilde{K} W
u_\alpha(l,m,\vec{q},t)~=~Hu_\alpha(l,m,\vec{q},t), \nonumber \\
\end{eqnarray}
with
\begin{eqnarray}
\label{eq:hamilton} &&H_{l,l^\prime}(m) =q(W^T \tilde{K}
W)_{l,l^\prime}  \\
&=& q \left(a_{lm} \delta_{l+1,l^\prime} +a_{l-1,m}
\delta_{l-1,l^\prime} \right)
\left(v^\ast_F-\frac{k^2_F}{(2\pi)^3}f_F(l,l)\right)^{1/2}\left(v^\ast_F-\frac{k^2_F}{(2\pi)^3}f_F(l^{\prime},l^{\prime
})\right)^{1/2}. \nn
\end{eqnarray}
It is apparent that the Hamiltonian in Eq.~(\ref{eq:hamilton})
is hermitian, i.e.,
$H=H^\dagger$.

\begin{figure}[htb]
\begin{center}
\includegraphics[width=0.6\textwidth]{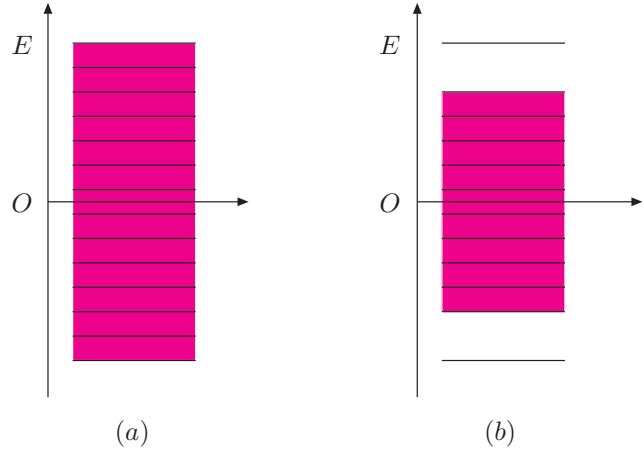}
\end{center}
\caption{Energy levels of Fermi liquid. $(a)$ represents the case of $f_F(l,l)=0$, where only continuous energy levels of nucleon-hole pairs are excited, and $(b)$ stands for the case of $f_F(l,l)>0$, where
two discrete energy levels of the collective excitation of nuclear matter appear except for continuous energy levels.}
\label{fig:1909251109}
\end{figure}

In Ref.~\cite{Wen}, Fermi liquid function is expanded in the complex Fourier series in the 2-dimensional momentum space. When the collective excitation of nuclear matter is studied, it is reasonable to expand the quasi-nucleon density in spherical harmonics in the 3-dimensional momentum space.
Meanwhile, if the direct interaction of nucleons is eliminated and only the exchange interaction of nucleons is taken into account, Fermi liquid function can be expanded in spherical harmonics. According to Eq. (\ref{eq:1909250746}), the exchange potential between two nucleons near Fermi surface is written as
\ba
\label{eq:202008021705}
&&V_{eff}(\vec{k}_1-\vec{k}_2)=\sum_{l_1=0}^{+\infty} \sum_{m_1=-l_1}^{l_1}
\sum_{l_2=0}^{+\infty} \sum_{m_2=-l_2}^{l_2} f_F(l_1,l_2) \delta_{l_1,l_2} \delta_{m_1,m_2}
Y^\ast_{l_1,m_1}(\theta,\phi)~Y_{l_2,m_2}(\theta^{\prime},\phi^{\prime}) \nn \\
&=&\sum_{l_1=0}^{+\infty} \sum_{m_1=-l_1}^{l_1}
f_F(l_1,l_2) Y^\ast_{l_1,m_1}(\theta,\phi)~Y_{l_1,m_1}(\theta^{\prime},\phi^{\prime}) \nn \\
&=&\sum_{l_1=0}^{+\infty} f_F(l_1,l_1) P_{l_1}(\cos \xi)~\frac{2{l_1}+1}{4\pi},
\ea
with $\xi$ the angle between two momenta $\vec{k}_1$ and $\vec{k}_2$. In order to obtain Eq.~(\ref{eq:202008021705}), the equation
$$P_l(\cos \xi)=\frac{4\pi}{2l+1}\sum_{m=-l}^{l} Y^\ast_{l,m}(\theta,\phi)~Y_{l,m}(\theta^{\prime},\phi^{\prime})$$ is used.
At this point, it is consistent with Eq.~(11) in Ref.~\cite{Friman:2019ncm}, where the interaction between two nucleons is expanded in Legendre polynomials.

If the ground state of a nucleus has spin and parity $J^p=0^+$, hence the excited nucleon-hole state after the electric dipole transition must have the quantum number $J^p=1^-$.
For most of doubly even nuclei,  their first excited states always have quantum numbers $J^p=2^+$, however,   doubly magic nuclei, such as $^{16}O$, $^{40}Ca$ and $^{208}Pb$, have a lower lying $J^p=3^-$ state\cite{Povh}.
The quasi-nucleon density and Fermi liquid function can be expanded in spherical harmonics in the momentum space, and the quantum number $l=0,1,2,...$. Actually, $l$ is not the quantum number of the orbital angular momentum, but corresponds to the spin of nuclei.

Assuming Fermi liquid function is zero, i.e., $f_F(l,l)=0$,
continuous energy levels of the Hamiltonian in
Eq.~(\ref{eq:hamilton}) are produced, which correspond to
nucleon-hole excitations in nuclear matter.
However, if the value of Fermi liquid function is
large enough, and $f_F(l,l)>0$, it is
possible to produce two discrete energy levels other than the
continuous ones. The positive discrete energy level represents the creation of
the collective excitation mode, while the negative
one stands for the annihilation of the collective excitation mode of
nuclear matter, as depicted in Fig.~\ref{fig:1909251109}.

In this work, when the collective excitation with determined $l$ is studied, Fermi liquid functions of other $l$ values is chosen to be zero, only $f_F(l,l)$ with determined $l$ is conserved in the Hamiltonian. It is found that the discrete excitation energy decreases with the $l$ value increasing. When $l$ is large enough, the discrete excitation energy level becomes identical and indistinguishable to the continuous levels, thus only the lower $l$ excitation modes are taken into account in the calculation.

\section{The collective excitation of nuclear matter}
\label{sect:Evskf}

In the relativistic mean-field approximation, the parameters are fitted
according to the saturation properties of nuclear matter. However,
if the nonlinear self-interacting terms of scalar mesons are included
in the Lagrangian, the compression modulus of nuclear matter
will be in a reasonable range, and the equation of state of nuclear matter will not be too
stiff.
In this work, the parameter set NL3 is adopted in the calculation, i.e.,
$g_\sigma=10.217$, $g_\omega=12.868$, $g_2=-10.431fm^{-1}$,
$g_3=-28.885$, $m_\sigma=508.194$MeV, $m_\omega=782.501$MeV and
$M_N=939$MeV\cite{NL3}.

According to Eq.~(\ref{eq:velocity}), Fermi velocity is related to
the nucleon effective mass $M^\ast_N$, which takes form of
$M^\ast_N=M_N+g_\sigma \sigma_0$ in the relativistic mean-field
approximation, where $\sigma_0$ is the expectation of the scalar meson field
in nuclear matter. Apparently, the self-consistency must be conserved when
the calculation is performed.
Since the nucleon near Fermi surface is easier to be excited
than the others, we assume $q\equiv k_F$ in Eq.~(\ref{eq:hamilton}),
and Fermi momentum $k_F=$1.36fm$^{-1}$ for the saturation nuclear matter.
In nuclei, there are a finite number of nucleons, which fill single-nucleon energy levels up to Fermi momentum $k_F$. In the momentum space, there is a Fermi ball with the radius of $k_F$, and the collective excitation of nuclei corresponds to the displacement of Fermi surface from the ball in the momentum space. In principle, the giant resonances of finite nuclei can also be analyzed in Fermi liquid model.

\subsection{The case of $m=0$}

As mentioned above, $l$ represents the spin of the nucleus, thus $m$ stands for the spin orientation of the nucleus. Both $l$ and $m$ are good quantum numbers.
In Ref.~\cite{Friman:2019ncm}, the interaction between nucleons is expanded in Legendre polynomials, and Legendre polynomial $P_l$ corresponds to spherical harmonic $Y_{l,0}$. Therefore, the case of $m=0$ is compared to the experimental data.
In this subsection, the quantum number $m$ in the
coefficient $a_{l,m}$ in Eq.~(\ref{eq:1903171236}) is assumed
to be zero, and then the collective excitation mode with $m=0$ is studied firstly.

\begin{figure}[htb]
\begin{center}
\includegraphics[width=0.6\textwidth]{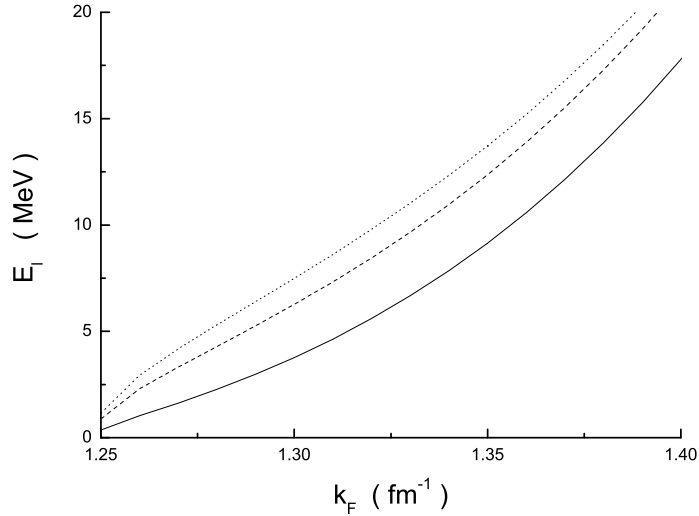}
\end{center}
\caption{Collective excitation energies of nuclear matter for different $l$ values .vs. Fermi momentum. The dotted line represents the case of $l=0$, the dashed line denotes the case of $l=1$, and the solid line stands for the case of $l=2$.}
\label{fig:e-kf}
\end{figure}

The collective excitation energy of nuclear matter $E_l$ with different $l$ values as functions of Fermi momentum $k_F$ are shown in Fig.~\ref{fig:e-kf}, where the dotted line represents the case of $l=0$, the dashed line denotes the case of $l=1$, and the solid line stands for the case of $l=2$.
It indicates that the collective excitation energy of nuclear matter $E_l$ increases with Fermi momentum $k_F$.

The experimental central energies of isoscalar and isovector giant monopole resonance(l=0), giant dipole resonance(l=1) and giant quadrupole resonance(l=2) for different nuclei are summarized in Table~\ref{tab:e-l}\cite{Youngblood,Berman,Woude}. It can be found that the energy of isovector giant resonances $E_V$ is about twice of that of isoscalar giant resonances $E_S$ respectively except for the $l=1$ case, which will be discussed detailedly in follows. Moreover, the collective excitation energy of heavy nuclei is less than the corresponding value of light nuclei.

The collective excitation energies of nuclear matter at the saturation density for $l=0,1,2$ are also listed in Table~\ref{tab:e-l}.
When the self-interacting terms of scalar mesons are taken into account, by solving the equation of motion of Fermi liquid in Eq.~(\ref{eq:Schrodinger}), the reasonable collective excitation energies of the saturation nuclear matter can be obtained, which are closer to the corresponding collective excitation energies of the nucleus ${}^{208}Pb$ than light nuclei.

If the self-interacting terms of scalar mesons are not included in the calculation of the relativistic mean-field approximation, the compression modulus of nuclear matter at the saturation density is about 500MeV, which is too large and nuclear matter becomes stiff. When the equation of motion in Eq.~(\ref{eq:Schrodinger}) is solved, the collective excitation energies are far lager than the corresponding experimental values, just as done in Ref.~\cite{Sun2010,Sun2012}.

\begin{table}[hbt]
\begin{center}
\begin{tabular}{ccccc}
\hline
  $l$      &       & $0$     & $1$     &  $2$ \\
\hline
$Theory$   & $E_l$ & $15.20$         & $13.88$     & $10.58$ \\
\hline
$^{208}Pb$ & $E_S$ & $14.17\pm0.28$  & $22.5$      & $10.9\pm0.1$ \\
           & $E_V$ & $26.0\pm3.0$    & $13.5\pm0.2$& $22$  \\
\hline
$^{144}Sm$ & $E_S$ & $15.39\pm0.28$  & $-$         & $-$  \\
\hline
$^{116}Sn$ & $E_S$ & $16.07\pm0.12$  & $-$         & $-$  \\
\hline
$^{90}Zr$  & $E_S$ & $17.89\pm0.20$  & $-$         & $14.41\pm0.1$ \\
           & $E_V$ & $28.5\pm2.6$    & $16.5\pm0.2$& $-$ \\
\hline
$^{40}Ca$  & $E_S$ & $-$             & $-$         & $17.8\pm0.3$ \\
           & $E_V$ & $31.1\pm2.2$    & $19.8\pm0.5$& $32.5\pm1.5$ \\
\hline
\end{tabular}
\caption{The collective excitation energies of the saturation nuclear matter $E_l$ for $l=0,1,2$ and the corresponding experimental central energies of the isoscalar and isovector giant resonances of nuclei with different $l$, which are labeled as $E_S$ and $E_V$, respectively. All energies are in units of MeV. }
\label{tab:e-l}
\end{center}
\end{table}

Since the exchange term in Fermi liquid function is nonzero when the isospin and spin of two nucleons are the same as each other,  nucleons with spin up and spin down can oscillate either in phase ( S=0 spin scalar mode) or out of phase ( S=1 spin vector mode), Moreover , protons and neutrons can oscillate either in phase (T=0 isoscalar mode) or oppositely (T=1 isovector modes). Altogether, there are four kinds of collective excitation modes in nuclear matter\cite{Ning}.

The label $\alpha$ in Eq.~(\ref{eq:Schrodinger}) represents the isospin and spin of nucleons, so the $E_l$ values listed in Table~\ref{tab:e-l} correspond to the isoscalar and spin scalar modes of nuclei(T=0, S=0), respectively.

In symmetric nuclear matter, where the number density of protons equals to that of neutrons, the collective excitation energy of the isovector and spin scalar modes(T=1, S=0) is the sum of the excitation energies of protons and neutrons, thus it is about twice of the energy of the isoscalar and spin scalar mode(T=0, S=0).
In Table~\ref{tab:e-l}, the experimental central energies of the isovector giant resonance of $^{208}Pb$ and $^{90}Zr$ with $l=0$, and $^{40}Ca$ with $l=2$ are twice of the corresponding central energies of the isoscalar giant resonance, respectively. Thus it manifests the relation between the isovector and spin scalar mode and the isoscalar and spin scalar mode is correct except for the $l=1$ case, which will be discussed in follows.

Similarly, the above analysis can be used to the isoscalar and spin vector modes(T=0, S=1), whose energy is also about twice of the energy of the isoscalar and spin scalar mode(T=0, S=0). Especially, it should be emphasized that the collective excitation energy of the isovector and spin vector mode (T=1,S=1) might be  about four times of the energy of of the isoscalar and spin scalar mode(T=0, S=0), which need to be examined by experiments in future.

The relation mentioned above is not available for the giant dipole resonance.
The dipole deformation of the nucleus is really a shift of the
center of mass. In the $\gamma-$induced reaction, the isovector giant dipole resonance of
the nucleus actually corresponds to the collective excitation of protons, and neutrons move on the opposite direction in the center of mass system of the nucleus.
The isoscalar giant
dipole resonance in $^{208}Pb$ with a central energy at
$E=22.5MeV$, using the $(\alpha,\alpha^\prime)$ cross sections at
forward angles\cite{Sun2010}, might
correspond to the spin vector mode with $l=1$, where nucleons with spin up and spin down oscillate out of phase. Therefore, the central energy is about twice of that of the dipole deformation mode.

\subsection{The case of $m \neq 0$}

The Hamiltonian in Eq.~(\ref{eq:hamilton}) is not only relevant to quantum number $l$, but to quantum number $m$. In this subsection, the case of
$m \neq 0$ will be discussed.

\begin{table}[hbt]
\begin{center}
\begin{tabular}{c|ccccccc}
\hline
 $E_{l,m}~(MeV)$ & $m=-3$ & $m=-2$ & $m=-1$ & $m=0$ & $m=1$ & $m=2$ & $m=3$\\
\hline
$l=0$ &         &         &         & $15.20$&        &         & \\
$l=1$ &         &         & $10.12$ & $13.88$& $10.12$&         &  \\
$l=2$ &         &$10.09$  & $10.11$ & $10.58$& $10.11$& $10.09$ & \\
$l=3$ & $10.06$ &$10.09$  & $10.11$ & $10.15$& $10.11$& $10.09$ &$10.06$ \\
\hline
\end{tabular}
\caption{Collective excitation energies of the saturation nuclear matter for different values of $l$ and $m$.}
\label{tab:e-lm}
\end{center}
\end{table}

The collective excitation energies of the saturation nuclear matter with Fermi momentum $k_F=$1.36fm$^{-1}$ for different $l$ and $m$ are listed in Table~\ref{tab:e-lm}. For the same value of $l$, $m$ changes from $-l$ to $l$, and the collective excitation energy with $m$ is the same as the minus $m$ case.
When the $l$ value is conserved, the collective excitation energy of the saturation nuclear matter decreases with the absolute value of $m$ increasing.

The collective excitation energies for nonzero $m$ are similar to that of the $m=0$ case. This is because the collective excitation energies of nuclear matter are not calculated from a microscopic nuclear theory, but treated as eigenvalues of Hamiltonian in Eq.~(\ref{eq:hamilton}), which is obtained through a bosonization of Boltzmann equation. Therefore, the results are only qualitative and more accurate research work on this topic is expected.

\section{Conclusions}
\label{sect:conclusion}

Landau Fermi liquid model is bosonized and used to describe the collective fluctuation of the two-dimensional Fermi system in Ref.~\cite{Wen}. In this work, this method is extended to the three-dimension Fermi system, and the isospin and spin of the fermion are also taken into account. Therefore, we tried to study the collective excitation of nuclear matter in the bosonized Fermi liquid model.

The equation of state of nuclear matter can be obtained reasonable in the framework of Walecka model, where the nonlinear self-interacting terms of scalar mesons are included in order to produce a correct compression modulus of nuclear mater in the relativistic mean-field approximation. Otherwise, the equation of state of nuclear matter would be stiff and the averaged energy per nucleon would be too large at high densities.
When the nonlinear self-interacting terms of scalar mesons are included in the Lagrangian,
the collective excitation energies for different $l$ and $m$ can be obtained self-consistently, where $l$ and $m$ correspond to the total spin and the orientation of the total spin of the nucleus, respectively, and they are both good quantum numbers in nuclei. Moreover, the results with $m=0$ are compared with the central energies of giant resonances of finite nuclei, and it shows that they
are consistent with the corresponding central energies of the giant resonances of $^{208}Pb$, respectively.
Comparing to the results in Ref.~\cite{Sun2010} calculated ten years ago, the nucleon effective mass only changes with the Fermi momentum in nuclear matter, and will not be a parameter in the calculation. In this model, only the coupling constants in Walecka model are parameters, and the NL3 parameter set is adopted in this work. Actually, it is found that almost all parameter sets with the nonlinear self-interacting terms of scalar mesons, such as NL1 and NL-SH parameters\cite{NL3}, can lead to correct collective excitation energies at the saturation density of nuclear matter.

In Fermi liquid model, the collective excitations of nucleons with spin up and spin down are independence of each other, they can oscillate either in phase or oppositely, Similarly, it can be extended to the isospin case. Protons can oscillate either in phase or out of phase with neutrons in nuclear matter. Altogether, there are four kinds of collective excitation modes existing in nuclear matter, which are the isoscalar and spin scalar mode(T=0, S=0), the isoscalar and spin vector mode(T=0, S=1), the isovector and spin scalar mode(T=1, S=0), and the isovector and spin vector mode(T=1, S=1), respectively. The collective excitation energies of them are discussed according to the corresponding experimental data of finite nuclei, and it is found that the collective excitation energies of the isoscalar and spin vector mode(T=0, S=1) and the isovector and spin scalar mode(T=1, S=0) are about twice of those of the isoscalar and spin scalar mode(T=0, S=0), respectively. Moreover, it is predicted that the collective excitation energies of the isovector and spin vector mode(T=1, S=1) is about four times of those of the isoscalar and spin scalar mode(T=0, S=0) in nuclear matter. Certainly, it should be examined by the experiments in future.

Since the nuclear quasi-nucleon density and Fermi liquid function are both expanded in spherical harmonics, the collective excitation of nuclear matter with $m \neq 0$ is taken into account naturally.
For the case of $m \neq 0$, the collective excitation energy with fixed $(l,m)$ takes the same value as that of $(l,-m)$.
It is found that
the collective excitation energy of nuclear matter decreases with the absolute value of $m$ increasing when the $l$ value is fixed.
Especially, it should be emphasized that the exchange interaction between nucleons near Fermi surface is important to excite the nucleon-hole pair in nuclear matter and plays a critical role in the collective excitation of nuclear matter.
In the future work, the pion and $\rho$ meson exchange between nucleons will be taken into account, and their influence on the collective excitation of nuclear matter will be studied.
However, our model only supplies some qualitative results, and more precise and quantitative calculations with other models are expected.

\section*{Appendix: Fermi liquid function}

If the effective potential between two nucleons is $V_{eff}(\vec{r}^\pr-\vec{r})$,
the two-body interaction operator can be written as
\ba \label{eq:1901211103} \hat{V}&=&
\sum_{\alpha,\alpha^\pr, \beta, \beta^\pr} \int d^3 r \int d^3 r^\pr
\psi_{\alpha}^\dagger(\vec{r})
\psi_{\alpha^\pr}^\dagger(\vec{r}^{\pr})
V_{eff}
\left(\vec{r}^\pr-\vec{r} \right)
\psi_{\beta^\pr}(\vec{r}^{\pr}) \psi_{\beta}(\vec{r}), \ea
where
\ba \label{eq:1901231718} \psi^\dagger_{\lambda}(\vec{r})&=&
\frac{1}{(2\pi)^{3/2}} \int d^3 p b^\dagger(p,\lambda)
\bar{U}(p,\lambda) \gamma_0 \exp(i p \cdot r),
\ea
and
\ba
\label{eq:190909251749}
\psi_{\lambda}(\vec{r})&=& \frac{1}{(2\pi)^{3/2}} \int d^3 p
b(p,\lambda) \bar{U}(p,\lambda) \exp(-i p \cdot r), \ea
with $p$ the nucleon momentum in Minkowski space. Since there are not anti-nucleons
in the ground state of nuclear matter, the terms related to the creation
and annihilation operators of antinucleons are eliminated in
Eqs.~(\ref{eq:1901231718}) and (\ref{eq:190909251749}).

The scattering matrix element between two nucleons takes the form of
\be
\label{eq:1909251805}
S_{fi}^{(2)}=\la f |S^{(2)}| i
\ra
 =\la p^\pr \lambda^\pr,k^\pr \delta^\pr |S^{(2)}|
 p \lambda,k \delta \ra,
\ee
with
\be S^{(2)}=-i \int d t \hat{V}. \ee
The anti-symmetric wave function of the two-nucleon system can be written as
\be
\label{eq:1901211215} | p \lambda,k \delta \ra=\frac{1}{\sqrt{2}}
\left[|p \lambda \ra_1 |k \delta \ra_2 - |k \delta \ra_1 |p \lambda
\ra_2 \right], \ee and \be \label{eq:1901211216} \la p^\pr
\lambda^\pr,k^\pr \delta^\pr |=\frac{1}{\sqrt{2}} \left[{}_2\la
k^\pr \delta^\pr | {}_1\la p^\pr \lambda^\pr | -  {}_2\la p^\pr
\lambda^\pr |{}_1\la k^\pr \delta^\pr | \right], \ee
where the labels $1$ and $2$ represent the first and second nucleons,
respectively. In what follows, these two labels will be neglected.

In the non-relativistic approximation, Assuming the interaction between
two nucleons is realized instantaneously, we can make an approximation ${r^\pr}^0=r^0\rightarrow t$
in Eq.~(\ref{eq:1909251805}).
Moreover, the nucleon wave function
is independent on the momentum, and only relevant to the nucleon spin,
so we can obtain
 \be \label{eq:1901241057}
\bar{U}(k^\prime,\delta^\prime) U(k,\delta)
=\bar{U}(\delta^\prime) U(\delta) \nn \\
=\delta_{\delta^\prime,\delta}, \ee and
 \ba
\label{eq:1901231853}
\bar{U}(k^\prime,\delta^\prime)\gamma_\mu U(k,\delta)
%
%
~=~ \left\{\begin{array}{cc}
 \delta_{\delta^\prime,\delta}, & \mu=0, \\
 0, & \mu=1,2,3.
       \end{array}
\right.
\ea
If the effective potential $V_{eff}(\vec{q})$ is only relevant to the
exchanged momentum squared $\vec{q}^2$, and independent on the nucleon
spin, the scattering matrix element $S_{fi}^{(2)}$ can be written as
\ba \label{eq:1901231907}
S_{fi}^{(2)}&=&-i~\frac{1}{(2\pi)^2}
~\delta^{(4)}(k+p-{k^\pr}-{p^\pr})
\nn \\
&&\left[ V_{eff} \left(\vec{k}-\vec{k}^\pr \right)
 \delta_{\delta^\pr,\delta}~\delta_{\lambda,\lambda^\pr}
-V_{eff} \left(\vec{p}-\vec{k}^\pr \right)
\delta_{\delta^\pr,\lambda} \delta_{\lambda^\pr,\delta}\right].
\ea

The Lagrangian density of nuclear matter in Walecka model can be written as\cite{Walecka}
\begin{eqnarray}
\label{eq:201812021304}
 {\cal L}~&=&~\bar\psi
\left(i\gamma_{\mu}\partial^{\mu} -
M_N\right)\psi~+~\frac{1}{2}\partial_\mu\sigma\partial^\mu\sigma-\frac{1}{2}
m^2_\sigma \sigma^2_{}-\frac{1}{3}g_2 \sigma^3 -\frac{1}{4} g_3 \sigma^4
 -\frac{1}{4}\omega_{\mu\nu}\omega^{\mu\nu}+
\frac{1}{2}
m^2_\omega\omega_\mu\omega^\mu \nonumber \\
&&-g_\sigma\bar\psi\sigma\psi-g_\omega\bar\psi \gamma_\mu \omega^\mu
\psi,
\end{eqnarray}
where the tensor of the vector meson is
$\omega_{\mu\nu}~=~\partial_\mu\omega_\nu-\partial_\nu\omega_\mu$, and
$M_N$, $m_\sigma$ and $m_\omega$ represent masses of the nucleon,
$\sigma$ and $\omega$ mesons, $g_2$ and $g_3$ are coefficients related to the nonlinear seif-interacting terms of scalar mesons,
$g_\sigma$ and $g_\omega$ denote coupling constants of the nucleon to
$\sigma$ and $\omega$ mesons, respectively.

\begin{figure*}
\includegraphics{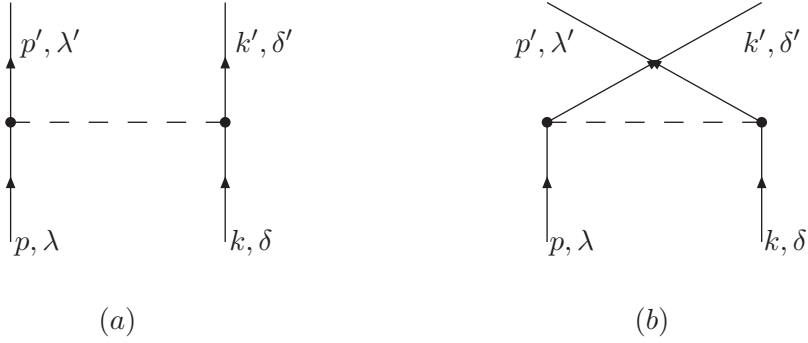}
\caption{\label{fig:181203}
Nucleon-nucleon interaction, where $(a)$ represents the direct interaction, and $(b)$ stands for
the exchange interaction.}
\end{figure*}

According to the Lagrangian density in Eq.~(\ref{eq:201812021304}),
the interacting Hamiltonian can be written as
\be
{\cal H_I}(x)=g_\sigma\bar\psi\sigma\psi+g_\omega\bar\psi \gamma_\mu
\omega^\mu \psi. \ee
Therefore, the second-order scattering matrix element takes the form of
\begin{equation}
\label{eq:S2} \hat{S}_2 ~=~\frac{(-i)^2}{2 !}\int d^{4}x_1 \int
d^{4}x_2
 T\left[{\cal H}_I(x_1){\cal H}_I(x_2) \right].
\end{equation}
In order to obtain the scattering amplitude of two nucleons, the Feynmann diagrams in
Fig.~\ref{fig:181203} must be calculated.

Firstly, the case that one vector meson exchanges between two nucleons
is studied. The scattering matrix element can be written as
\ba \label{eq:1901161201}
S^{(2)}_{fi}(\omega)&=&-g^2_\omega (2\pi)^4 \delta^4
(p^\prime+k^\prime-p-k)
\left(\frac{1}{(2\pi)^{3/2}} \right)^4 \nn \\
&&\left[\bar{U}(k^\prime,\delta^\prime)\gamma_\mu U(k,\delta)
\frac{-ig^{\mu \nu}}{(k^\prime-k)^2-m_\omega^2+i\varepsilon}
\bar{U}(p^\prime,\lambda^\prime)\gamma_\nu U(p,\lambda)
 \right. \nn \\
&&-\left. \bar{U}(k^\prime,\delta^\prime)\gamma_\nu U(p,\lambda)
\frac{-ig^{\mu \nu}}{(k^\prime-p)^2-m_\omega^2+i\varepsilon}
 \bar{U}(p^\prime,\lambda^\prime)\gamma_\mu U(k,\delta)
 \right]. \nn \\
\ea

The three-momentum of the nucleon is far lower than the
nucleon mass,
$|\vec{k}|<<M_N$, $|\vec{k}^\prime|<<M_N$, $|\vec{p}|<<M_N$,
$|\vec{p}^\prime|<<M_N$, which corresponds to the nucleon mass becomes infinite
 in the non-relativistic approximation, i.e., $M_N \rightarrow
+\infty$. Thus the zero component of the momentum of the intermediate vector
meson in the direct interaction of nucleons can be written as
 \ba
q_0&=&k^\prime_0-k_0 \nn \\
&=&\sqrt{\vec{k^\prime}^2+M_N^2}-\sqrt{\vec{k}^2+M_N^2} \nn \\
&\approx& M_N \left(1+\frac{\vec{k^\prime}^2}{2M_N^2} \right)
        - M_N \left(1+\frac{\vec{k}^2}{2M_N^2} \right) \nn \\
&=&\frac{\vec{k^\prime}^2}{2M_N}-\frac{\vec{k}^2}{2M_N} \nn \\
&\rightarrow&0. \ea
Similarly, it can be certified that the zero component of
the vector meson momentum in the exchanging interaction of nucleons $q_0^\prime=k_0^\prime-p_0$ tends
to zero in the non-relativistic approximation.

According to Eq.~(\ref{eq:1901231853}), the scattering matrix element
in Eq.~(\ref{eq:1901161201}) can be simplified as
\ba
\label{eq:1901191642} S^{(2)}_{fi}(\omega)&=&-g^2_\omega (2\pi)^4
\delta^4 (p^\prime+k^\prime-p-k)
\left(\frac{1}{(2\pi)^{3/2}} \right)^4 \nn \\
&&\left[\delta_{\delta^\prime,\delta} \delta_{\mu,0}
\frac{-ig^{\mu \nu}}{(k^\prime-k)^2-m_\omega^2+i\varepsilon}
\delta_{\lambda^\prime,\lambda} \delta_{\nu,0}
 \right. \nn \\
&&-\left. \delta_{\delta^\prime,\lambda} \delta_{\nu,0}
\frac{-ig^{\mu \nu}}{(k^\prime-p)^2-m_\omega^2+i\varepsilon}
 \delta_{\lambda^\prime,\delta} \delta_{\mu,0}
 \right]. \nn \\
&=&-g^2_\omega \frac{i}{(2\pi)^2}
\delta^4 (p^\prime+k^\prime-p-k) \nn \\
&&\left[
\frac{1}{(\vec{k}^\prime-\vec{k})^2+m_\omega^2+i\varepsilon}
 \delta_{\delta^\prime,\delta} \delta_{\lambda^\prime,\lambda}
 \right. \nn \\
&&-\left.
\frac{1}{(\vec{k}^\prime-\vec{p})^2+m_\omega^2+i\varepsilon}
\delta_{\delta^\prime,\lambda} \delta_{\lambda^\prime,\delta}
 \right]. \nn \\
\ea

Comparing Eqs.~(\ref{eq:1901231907}) with (\ref{eq:1901191642}), the
nucleon potential by exchanging a vector meson is written as
 \be
\label{eq:1901241814}
V^{\omega}_{eff}(\vec{q})=\frac{g_\omega^2}{\vec{q}^2+m_\omega^2}.
\ee
The nucleon potential by exchanging a scalar meson can be obtained similarly
\be
\label{eq:1901241815}
V^{\sigma}_{eff}(\vec{q})=\frac{-g_\sigma^2}{\vec{q}^2+m_\sigma^2}.
\ee
Therefore,
the total potential between two nucleons is the sum of potentials in Eqs.~(\ref{eq:1901241814}) and (\ref{eq:1901241815}),
 \be
\label{eq:1901241832}
V_{eff}(\vec{q})=\frac{-g_\sigma^2}{\vec{q}^2+m_\sigma^2}
+\frac{g_\omega^2}{\vec{q}^2+m_\omega^2}. \ee

In the coordinate representation, the nucleon potential in
Eq.~(\ref{eq:1901241832}) takes the form of Yukawa potential,
\ba \tilde{V}_{eff}(r)&=&\frac{1}{(2\pi)^3} \int d^3 q
V_{eff}(\vec{q})
\exp \left(i\vec{q} \cdot \vec{r} \right) \nn \\
&=&-\frac{g_\sigma^2}{4\pi} \frac{\exp(-m_\sigma r)}{r}
+\frac{g_\omega^2}{4\pi} \frac{\exp(-m_\omega r)}{r},
\ea
which is a
Fourier transformation of the potential $V_{eff}(\vec{q})$
in Eq.~(\ref{eq:1901241832}).

Fermi liquid function denotes the correlation between two nucleons near Fermi surface, as depicted in Fig.~\ref{fig:181203}. If the momentum, isospin and spin of outgoing nucleons are all the same as those of incoming nucleons, respectively, i.e., $p^\pr=p$, $k^\pr=k$, $\lambda^\pr=\lambda$, $\delta^\pr=\delta$,
Fermi liquid function is defined
as the sum of the direct and exchanging potential. According to the nucleon potential in
Eq.~(\ref{eq:1901241832}), Fermi liquid function can be written as
\be \label{eq:1901241841} f(\vec{p},\lambda;\vec{k},\delta)
=V_{eff}(0)-V_{eff}(\vec{p}-\vec{k})\delta_{\lambda,\delta}, \ee
which is consistent with the scattering matrix element of two nucleons in Eq.~(\ref{eq:1901231907}) in a non-relativistic approximation.

The first term in Fermi liquid function is a constant in the momentum space, which is relevant to the direct interaction of nucleons. Actually, the direct interaction of nucleons supplies a mean field in nuclear matter, and Fermi energy and velocity of nucleons change in the relativistic mean-field approximation, as shown in Eqs.~(\ref{eq:202007311120}) and (\ref{eq:velocity}). Only the second term in Fermi liquid function, which corresponds to the exchange interaction of nucleons, gives a contribution to the nucleon-hole excitation in nuclear matter. Therefore, only the second term is left in Fermi liquid function when Eq.~(\ref{eq:Schrodinger}) is solved.

Although the nucleon effective mass is calculated in the relativistic mean field approximation, Fermi liquid function takes a non-relativistic form in the momentum space, and Landau Fermi liquid model adopted in this work is still non-relativistic.


\begin{thebibliography}{50}

\bibitem{Greiner}
W. Greiner and J. A. Maruhn, {\it Nuclear Models}, Springer-Verlag,
Berlin, 1996.

\bibitem{Zhang}
  I.~Hamamoto, H.~Sagawa and X.~Z.~Zhang,
  Phys.\ Rev.\  C {\bf 57}, R1064, (1998).

\bibitem{Ma}
  Z.~Y.~Ma, A.~Wandelt, N.~Van Giai, D.~Vretenar, P.~Ring and L.~G.~Cao,
  Nucl.\ Phys.\  A {\bf 703}, 222, (2002).

\bibitem{Ring09}
  J.~Daoutidis and P.~Ring,
  Phys.\ Rev.\  C {\bf 80}, 024309, (2009).

\bibitem{SchuckRing} P. Ring and P. Schuck, {\it The Nuclear Many-Body
Problem}, Springer-Verlag, Berlin, 2004.


\bibitem{matsui}
  T.~Matsui,
  Nucl.\ Phys.\  A {\bf 370}, 365, (1981).

\bibitem{Blaizot:1978ocy}
  J.~P.~Blaizot,
  Phys.\ Lett.\  {\bf 78B}, 367, (1978).

\bibitem{Holzwarth:1981zz}
  G.~Holzwarth and T.~Yukawa,
  Nucl.\ Phys.\ A {\bf 364}, 29, (1981).




\bibitem{Friman96}
  B.~Friman and M.~Rho,
  Nucl.\ Phys.\  A {\bf 606}, 303, (1996).



\bibitem{Song:2000cu}
  C.~Song,
  Phys.\ Rept.\  {\bf 347}, 289, (2001).

\bibitem{Kamerdzhiev:2003rd}
  S.~Kamerdzhiev, J.~Speth and G.~Tertychny,
  Phys.\ Rept.\  {\bf 393}, 1, (2004).

\bibitem{Holt:2006ii}
  J.~W.~Holt, G.~E.~Brown, J.~D.~Holt and T.~T.~S.~Kuo,
  Nucl.\ Phys.\ A {\bf 785}, 322, (2007).

\bibitem{Ebran:2012hi}
  J.-P.~Ebran, E.~Khan, T.~Niksic and D.~Vretenar,
  Phys.\ Rev.\ C {\bf 87}, 044307, (2013).

\bibitem{Kolomeitsev:2016zid}
  E.~E.~Kolomeitsev and D.~N.~Voskresensky,
  Eur.\ Phys.\ J.\ A {\bf 52}, 362, (2016).

\bibitem{Roepke:2017bad}
  G.~Roepke, D.~N.~Voskresensky, I.~A.~Kryukov, and D.~Blaschke,
  Nucl.\ Phys.\ A {\bf 970}, 224, (2018)

\bibitem{Holt:2017uuq}
  J.~W.~Holt, N.~Kaiser and T.~R.~Whitehead,
  Phys.\ Rev.\ C {\bf 97}, 054325, (2018).

\bibitem{Grasso:2018app}
  M.~Grasso, D.~Gambacurta and O.~Vasseur,
  Phys.\ Rev.\ C {\bf 98}, 051303, (2018).


\bibitem{GhazanfariMojarrad:2018acp}
  M.~Ghazanfari Mojarrad, N.~S.~Razavi and S.~Vaezzade,
  Nucl.\ Phys.\ A {\bf 980}, 51, (2018).

\bibitem{Friman:2019ncm}
  B.~Friman and W.~Weise,
  Phys.\ Rev.\ C {\bf 100}, 065807, (2019).




\bibitem{Wen}
X. G. Wen, {\it Quantum Field Theory of Many-Body Systems}, Oxford
University Press, Oxford, 2004.

\bibitem{Sun2010} B. X. Sun, arXiv:1003.1683[nucl-th] and references therein.
\bibitem{Sun2012} B. X. Sun, {\it Nuclear Structure in China 2010,
Proceedings of the 13th National Conference on Nuclear Structure
in China, 25-31 July, 2010, Chifeng, China},  p181-186, arXiv:1201.5723[nucl-th].

\bibitem{Mayugang}
H. L. Liu, Y. G. Ma, A. Bonasera, X. G. Deng, O. Lopez and M. Veselsky, Phys. Rev. C {\bf 96}, 064604,(2017).

\bibitem{Povh}
B. Povh, K. Rith, C. Scholz, F. Zetsche and W. Rodejohann, {\it Particle and Nuclei: An Introduction to the Physical Concepts}, 7th Edition, Springer-Verlag, Heidelberg, 2015.

\bibitem{NL3} G. A. Lalazissis, J. Koenig and P.Ring,
Phys. Rev. C {\bf 55} 540, (1997).

\bibitem{Youngblood}
  D.~H.~Youngblood, H.~L.~Clark and Y.~W.~Lui,
  Phys.\ Rev.\ Lett.\  {\bf 82}, 691, (1999).

\bibitem{Berman} B. L. Berman and S. C. Fultz, Rev. Mod. Phys. {\bf
47},  713, (1975).

\bibitem{Woude} A. Van de Woude, Prog. Part. Nucl. Phys. {\bf 18},
217, (1987).

\bibitem{Ning}
P. Z. Ning, L. Li, D. F. Min, {\it Fundamental Nuclear Physics: Nucleons and Nuclei} (in Chinese), Higher Education Press, Beijing, 2003.


\bibitem{Walecka}
  J. D. Walecka, {\it Theoretical Nuclear and Subnuclear Physics}, 2nd Edition, World Scientific
  Press, Singapore, 2004.

\end{thebibliography}
\end{document}